\documentclass[prc,aps,floatfix,groupedaddress,amsmath,amssymb]{revtex4}
\usepackage{graphicx}
\usepackage{epsfig}
\def\bea{\begin{eqnarray}}
\def\eea{\end{eqnarray}}
\def\be{\begin{equation}}
\def\ee{\end{equation}}

\def\P3{{\cal P}_t}
\def\J3{{\cal J}}
\def\T3{{\cal T}}

\def\beq{\begin{equation}}
\def\eeq{\end{equation}}
\def\ear{\end{array}}

\begin{document}

\title{From the crust to the core of Neutron Stars on a microscopic basis}

\author{M. Baldo, G. F. Burgio}

\affiliation{
INFN Sezione di Catania, and Dipartimento di Fisica e Astronomia,
Universit\'a di Catania, Via Santa Sofia 64, 95123 Catania, Italy}

\author{M. Centelles, B. K. Sharma and X. Vi\~nas}

\affiliation{Departament d'Estructura i Constituentes de la Mat\`eria and Institut de Ci\`encies del Cosmos,
Facultat de F\'{\i}sica, Universitat de Barcelona, Diagonal 645, 08028 Barcelona, Spain}
%\date{\today}

\begin{abstract}
Within a microscopic approach the structure of Neutron Stars is usually studied by modelling the homogeneous nuclear matter of the core by a suitable Equation of State, based on a many-body theory, and the crust by a functional based on a more phenomenological approach. We present the first calculation of Neutron Star overall structure by adopting for the core an Equation of State derived from the Brueckner-Hartree-Fock theory and for the crust, including the pasta phase, an Energy Density Functional based on the same Equation of State, and which is able to describe accurately the binding energy of nuclei throughout the mass table. Comparison with other approaches is discussed. The relevance of the crust Equation of state for the Neutron Star radius is particularly emphasised.

\end{abstract}


\maketitle

%===============================================================================
\section{Introduction}

A convergent effort of experimental and theoretical nuclear physics has been developing along several years to determine the structure and properties of Neutron Stars (NS). These studies are expected to reveal the properties and composition of neutron-rich nuclear matter at high density and at the same time the properties of exotic nuclei that are present in the crust and cannot be produced in laboratory because of the large asymmetry. The interpretation of the signals coming from the astrophysical observations on the processes and phenomena that occur in NS requires reliable theoretical inputs.
 The interplay between the observations and the theoretical predictions has stimulated an impressive progress in the field, and it is expected to help answering many fundamental questions on the properties of matter under extreme conditions and the corresponding elementary processes that can occur. Among others, we mention the maximum mass beyond which a NS collapses to a black hole, the baryon composition of matter at high density, and the properties of extremely asymmetric matter.
It is therefore of great interest to have a sound theoretical background for the development of the field, in order to reduce the uncertainty on the possible conclusions one can draw from these studies.
In particular it can be of great help to develop a unified theory which is able
to describe on a microscopic level the overall structure of NS, from the crust
to the inner core. This is not a simple task, since the methods developed for
homogeneous nuclear matter cannot be easily extended to nuclei and to the
non-homogeneous matter present in NS crust. Recently
\cite{BCPM,baldo08,baldo10}, an Energy Density Functional (EDF) to describe
finite nuclei has been developed, based on the nuclear matter EOS
derived from the Brueckner-Hartree-Fock (BHF) scheme. We employ this EOS and the
corresponding EDF to describe the whole NS structure and compare with the
results obtained with the few other methods that encompass the whole NS
structure. We compare also with the few semi-microscopic approaches, where the
crust and the core can be also described within the same theoretical scheme.
%\part
In this paper we limit the treatment to only nucleon degrees of freedom, neglecting the possibility of the appearance of exotic components like hyperons and quarks, for which the uncertainty is too large to perform a fruitful comparison with other approaches.
\par
The present article is organized as follows. In Sect.~II we introduce the models of the EOS considered in this work. Sect.~III is devoted to the description of our calculations of the NS crust, whereas in Sect.~IV we describe the calculation of the EOS of the NS core. In Sect.~V we discuss the results for the mass and radius of a NS. Finally, our concluding remarks are presented in Sect.~VI.

%===============================================================================
\section{EOS of nuclear matter}
\label{s:EOS}
In this section we first remind briefly the BHF method for the nuclear matter EOS.
This theoretical scheme is based on
the Brueckner--Bethe--Goldstone (BBG) many-body theory, which is the linked cluster 
expansion of the energy per nucleon of nuclear matter (see chapter 1 of Ref.~\cite{book} and references therein).
In this many--body approach one systematically replaces the bare nucleon-nucleon (NN) $v$ interaction by the Brueckner reaction
matrix $G$, which is the solution of the  Bethe--Goldstone equation 
\begin{equation}
G[\rho;\omega] = v  + \sum_{k_a k_b} v {{|k_a k_b\rangle  Q  \langle k_a k_b|}
  \over {\omega - e(k_a) - e(k_b) }} G[\rho;\omega], 
\end{equation}                                                           
\noindent
where $\rho$ is the nucleon number density, and $\omega$ the starting energy.
The single-particle energy $e(k)$,
\begin{equation}
e(k) = e(k;\rho) = {{\hbar^2 k^2}\over {2m}} + U(k;\rho),
\label{e:en}
\end{equation}
\noindent
and the Pauli operator $Q$ constrains the intermediate
baryon pairs to momenta above the Fermi momentum. The Brueckner--Hartree--Fock (BHF) approximation for the 
single-particle potential
$U(k;\rho)$  using the  {\it continuous choice} is
\begin{equation}
U(k;\rho) = {\rm Re} \sum _{k'\leq k_F} \langle k k'|G[\rho; e(k)+e(k')]|k k'\rangle_a,
\end{equation}
\noindent
where the subscript ``{\it a}'' indicates anti-symmetrization of the 
matrix element.  
Due to the occurrence of $U(k)$ in Eq.~(\ref{e:en}), these equations constitute
a coupled system that has to be solved in a self-consistent manner
for several  momenta of the particles involved, at the considered densities. 
In the BHF approximation the energy per nucleon is
\begin{equation}
{E \over{A}}  =  
          {{3}\over{5}}{{\hbar^2 k_F^2}\over {2m}}  + {{1}\over{2A}}
~ \sum_{k,k'\leq k_F} \langle k k'|G[\rho; e(k)+e(k')]|k k'\rangle_a. 
\end{equation}
\noindent
 The nuclear EOS can be
calculated with good accuracy in the Brueckner two hole-line 
approximation with the continuous choice for the single-particle
potential, since the results in this scheme are quite close to the 
calculations which include also the three hole-line
contribution \cite{song}. 
However, as it is well known, the  
non-relativistic calculations, based on purely two-body interactions, fail 
to reproduce the correct saturation point of symmetric nuclear matter and one needs
to introduce three-body forces (TBFs).
In our approach the TBF is reduced to a density dependent two-body force by
averaging over the position of the third particle. 
In this work we will illustrate results based on the so-called Urbana model \cite{uix}. 
The nuclear matter EOS was calculated in previous works \cite{BCPM} for both symmetric matter and neutron matter.
The TBF produce in symmetric matter a shift of the saturation point of about
$+1$ MeV in energy.
For computational purpose, on the calculated symmetric matter EOS an educated
polynomial fit was performed with a fine tuning of the two parameters
contained in the TBF, as in references \cite{bbb,zhou}, in order to get an optimal saturation point (the minimum), $E/A \,=\, -0.16$ MeV and $\rho_0 \,=\, 0.16$ fm$^{-3}$. The higher density  EOS, above $0.62$ fm$^{-3}$, is quite smooth and the calculated points have been interpolated numerically without any polynomial fit.\par
This EOS was used to construct an EDF  for describing finite nuclei, as
explained in ref. \cite{BCPM}. The nuclear matter EOS is assumed to be the bulk
part of the functional. One then needs few additional phenomenological
parameters to fit a large set of nuclear binding energies throughout the
nuclear mass table. The additional ingredients of the functional are a surface energy
term, necessary to describe correctly the nuclear surface, which is absent in
nuclear matter, the single-particle spin-orbit interaction, typical of finite
nuclei, and a pairing contribution to describe open-shell nuclei. The bulk
symmetry energy is directly taken from the nuclear matter EOS, by a quadratic
interpolation between pure neutron matter and symmetric nuclear matter. This
approximation is a good one, as it has been checked by comparing with
calculations in asymmetric nuclear matter. As explained in \cite{BCPM}, no explicit surface symmetry energy was introduced. With a set of three parameters it was possible to get an overall fit of the absolute nuclear binding of 579 even-even nuclei \cite{BCPM} of known experimental masses with an average quadratic deviation of 1.58 MeV. The deviation for the charge radii of 313 even-even nuclei \cite{BCPM}, which were not included in the fit, turns out to be of 0.027 fm. Both deviations compete with the best functionals present in the literature. This gives confidence to the use of this functional for the study of the NS crust, where very asymmetric nuclei appear. More details on this EDF, called BCPM (Barcelona-Catania-Paris-Madrid) hereafter, can be found in ref. \cite{BCPM}.
\par 
There are only few nuclear EOS that have been devised and used to cover the whole NS structure.
A partly phenomenological approach, based on the compressible liquid drop model, has been developed \cite{Lattimer} by Lattimer and Swesty (LS). It can cover the whole range of density, including the crust and the pasta phase, and it gives a complete description of the NS matter structure.
It also includes the temperature dependence of the EOS, but for our purpose we
will employ it only in the zero temperature limit. There are different versions
of this EOS, each one corresponding to a different incompressibility. This EOS is
derived from a macroscopic functional and it is compatible with an accurate mass
formula throughout the nuclear table. Therefore, it belongs to the set of EOS
within which we intend to make a comparison. \par
Another approach that has the same characteristics has been developed \cite{Shen} by Shen et al. (SH).
It is based on the relativistic mean field formalism and the Thomas-Fermi (TF)
scheme with trial densities. Also in this case we will employ this EOS in the zero temperature limit.
It does not include the pasta phase, but we think that this will not affect the
relevance of the comparison.
We will not describe in details the LS and SH
Equations of State, since they are available directly on the web 
in functional and tabular forms,
both for LS \cite{LSweb} and SH \cite{SHweb} cases.  
\par
Special mention must be given to the works 
where Skyrme forces are used to calculate 
the whole structure of NS.  
In ref. \cite{Haensel} the
force SLy4 \cite{Sly} was used, and since
this force was adjusted to
reproduce the pure neutron matter EOS of Friedman and
Pandharipande \cite{FP}, this approach
contains some microscopic input.
For the crust the compressible Liquid Drop
Model was used, with parameters
again extracted from the SLy4 force.
We will comment on this
approach in section V dedicated to the
results of the different EOS on the NS
structure.\par
More recently, other EOSs covering the whole NS have been derived by the Brussels-Montreal group \cite{BSK}. They use the Skyrme forces BSk19, BSk20, and BSk21 \cite{goriely10}, each one adjusted to the known masses of nuclei and constrained to reproduce a different microscopic neutron matter EOS with different stiffness at high density. The inner crust is treated in the extended Thomas-Fermi approach with trial densities including shell corrections for protons via the Strutinsky integral method. The basic physical characteristics of these EOS models of NS have been fitted by analytical expressions to facilitate their inclusion in astrophysical simulations \cite{potekhin13}. A comparison with the results of this approach is left for future study.
\section{The crust region}
In this Section, we report on the main results for the crust and pasta
phase calculated with the BCPM functional \cite{BCPM,baldo08,baldo10}, whose
bulk part is the EOS calculated in the BHF scheme, and we compare with other
approaches.
\par 
For the BHF case we use the corresponding EDF of refs. \cite{BCPM,baldo08,baldo10}. Our crust calculations are done in the Thomas-Fermi (TF) scheme which was developed in ref. \cite{sil02} for a general Skyrme functional, and which has been suitably adapted here for the present BCPM functional, as well as for the pasta phases. As in all approaches developed up to now, the crust will be described within the Wigner-Seitz approximation. Different shapes of the elementary cells of the lattice are tested in order to determine the most favorable structure.
The considered periodic structures include the spherical one, the layer or
slab (``lasagne''), and the cylinder or rod (``spaghetti''). In each case the Coulomb lattice
energy is then added, assuming a body-centered structure for the spherical case.

\subsection{The self-consistent Thomas-Fermi approach for the crust}
In this subsection we describe briefly the TF method that has been used to calculate the structure and the EOS of the crust on the basis of the BCPM energy density functional \cite{BCPM,baldo08,baldo10}. A complete report on the formalism and on the results will be presented elsewhere.
The total energy of an ensemble of neutrons, protons and electrons
in a Wigner-Seitz (WS) cell of volume $V_{c}$ is given by
\begin{eqnarray}
{E} = \int_{V_{c}} dV
\left[{\cal H}\left(\rho_{n},\rho_{p}\right) + {\epsilon}_{\rm el}
+ {\epsilon}_{\rm coul}+ {\epsilon}_{\rm ex} +m_{n}\rho_{n} + m_{p}\rho_{p}\right],
\label{eq1}
\end{eqnarray}
where ${\cal H}\left(\rho_{n},\rho_{p}\right)$ is the nuclear energy density
and $\rho_n$ and $\rho_p$ are the neutron and proton number densities, respectively. In the TF approach, ${\cal H}$ is given by
\begin{eqnarray}
{\cal H}\left(\rho_{n} ,\rho_{p}\right) = \frac{\hbar^{2}}{2m_{n}}\frac{3}{5}\left(3\pi^{2}\right)^{2/3}
\rho_{n}^{5/3}(\vec{r}) +
\frac{\hbar^{2}}{2m_{p}}\frac{3}{5}\left(3\pi^{2}\right)^{2/3}
\rho_{p}^{5/3}(\vec{r}) + {\cal
V}\left(\rho_{n}(\vec{r}),\rho_{p}(\vec{r})\right) .
\label{eq2}
\end{eqnarray}
The two first terms of Eq.~(\ref{eq2}) correspond to the neutron and proton TF
kinetic energy densities and ${\cal V}\left(\rho_{n},\rho_{p}\right)$ is the
interacting part produced by the
BCPM functional \cite{BCPM,baldo08,baldo10}.

The term ${\epsilon}_{\rm el}$ is the energy density due to the motion of the electrons. Since their Fermi energy is much higher than the Coulomb energy, one can approximate ${\epsilon}_{\rm el}$ by the energy density of a uniform relativistic gas, which is given by
\begin{eqnarray}
{\epsilon}_{\rm el}=
\frac{m_{e}^{4}c^{5}}{8\pi^{2}\hbar^{3}}\left[x_{e}\left(2x_{e}^{2}+1\right)
\sqrt{x_{e}^{2}+1}-\ln\left(x_{e}+\sqrt{x_{e}^{2}+1}\right)\right]
\label{eq3}
\end{eqnarray}
in terms of the dimensionless ratio $x_{e}={p_{Fe}}/{m_{e}c}$, where $p_{Fe}=\hbar c \left(3\pi^{2}\rho_{e}\right)^{1/3}$ is the Fermi momentum of the electrons, $m_{e}$ is the electron rest mass, and $\rho_e$ is the electron number density.

The term ${\epsilon}_{\rm coul}$ in Eq.~(\ref{eq1}) is the Coulomb energy density arising both from the direct part of the proton-proton and electron-electron interactions and from the proton-electron interaction. Under the assumption that the electrons are uniformly distributed in the WS cell, this term can be written as
\begin{eqnarray}
{\epsilon}_{\rm coul}=
\frac{1}{2} \left(\rho_{p}\left(\vec{r}\right)-\rho_{e}\right) \,
\left(V_{p}(\vec{r}) - V_{e}(\vec{r})\right)
\,=\,
\frac{1}{2} \left(\rho_{p}\left(\vec{r}\right)-\rho_{e}\right)
\int
\frac{e^{2}}{|\vec{r}-{\vec{r}\,'}|}\left(\rho_{p}({\vec{r}\,'})-\rho_{e}
\right)d{\vec{r}\,'}.
\label{eq4}
\end{eqnarray}
The exchange part of the proton-proton and electron-electron interactions
is calculated in the Slater approximation; its contribution is
\begin{equation}
{\epsilon}_{\rm ex} = -\frac{3}{4}\left(\frac{3}{\pi}\right)^{1/3}
\! e^2 \left({\rho_{p}}^{4/3}(\vec{r})+{\rho_{e}}^{4/3}\right) .
\end{equation}

At variance with previous calculations of TF type in the non-relativistic
framework \cite{oyamatsu93,gogelein07,onsi08,BSK}, where the proton and neutron
densities were parametrized and the minimization of Eq.~(\ref{eq1}) was
performed in a restricted variational approach,
we perform here a fully self-consistent variational calculation of the energy given by
Eq.~(\ref{eq1}). The minimization of the total energy is performed under the constraints of a given average density $\rho_B$ in the WS cell of size $R_c$ and of charge neutrality within the cell. Some self-consistent TF calculations in the inner crust using the Relativistic Mean Field Approach have been reported in the literature \cite{cheng97,avancini08,avancini09,avancini12}.

The variational Euler-Lagrangian equations of our problem are obtained
by taking functional derivatives with respect to the neutron, proton and
electron densities
in Eq.~(\ref{eq1}) including the aforementioned constraints. The ensuing
variational equations read
\begin{eqnarray}
\frac{\delta{\cal H}\left(\rho_{n},\rho_{p}\right)}{{\delta}{\rho_{n}}}
+ m_{n}- \mu_{n} = 0,
\label{eq5}
\end{eqnarray}
\begin{eqnarray}
\frac{\delta{\cal H}\left(\rho_{n},\rho_{p}\right)}{{\delta}{\rho_{p}}}
+ V_{p}\left(\vec{r}\right) - V_{e}(\vec{r})
-\left(\frac{3}{\pi}\right)^{1/3}\!\!\! e^2 {\rho_{p}}^{1/3} \left(\vec{r}\right)
+ m_{p}- \mu_{p} = 0,
\label{eq6}
\end{eqnarray}
\begin{eqnarray}
\sqrt{p_{Fe}^{2}+m_{e}^{2}}
- V_{p}\left(\vec{r}\right) + V_{e}(\vec{r})
-\left(\frac{3}{\pi}\right)^{1/3}\!\!\! e^2 {\rho_{e}}^{1/3} - \mu_{e} = 0,
\label{eq7}
\end{eqnarray}
together with the $\beta$-equilibrium condition
\begin{eqnarray}
\mu_{e}= m_{n}-m_{p}+\mu_{n}-\mu_{p}
\label{eq8}
\end{eqnarray}
which is imposed by the previously mentioned constraints applied to
Eq.~(\ref{eq1}).

For a given baryon density $\rho_B$ and an assumed size $R_c$ of the cell, the
set of Eqs.~(\ref{eq5})--(\ref{eq8}) is solved self-consistently following
the method described in Ref.~\cite{sil02} in order to find the composition (A,Z) of minimal energy
which corresponds to the prescribed $\rho_B$ and $R_c$ values and satisfies
$\beta$ equilibrium. Next, we perform the search of the optimal size of the cell
for the given baryon density $\rho_B$ by repeating the calculation for different
values of $R_c$. It must be pointed out that this calculation is a very delicate
task from the numerical point of view. The reason is that the minimum energy as
a function of $R_c$ is usually extremely flat and the differences of the total
energies involved are of the order of a fraction of a few keV and sometimes of
only a few eV.

The method of solving Eqs.~(\ref{eq5})--(\ref{eq8}) is not restricted to
spherical symmetry and it can be extended to WS cells with planar symmetry
(slabs) or cylindrical symmetry (rods). The calculations with these
non-spherical geometries are simplified if one considers slabs and rods of
infinite extension in the perpendicular direction to the size $R_c$. Although
the number of particles and the total energy of these cells is infinite, the
number of particles and energy per unit area (slabs) or unit length (rods) is
finite, and consequently the same is true for the total energy per baryon or per
unit of volume. With the choice of geometries that are infinite in the
perpendicular direction to the size $R_c$, the total energy per unit area or
unit length is easily derived from
Eq.~(\ref{eq1}). Taking $dV = Sdx$ (slabs) or $dV = {\pi}Lrdr$ (rods) reduces
the calculation of the total energy to a 1-dimensional or 2-dimensional integral
over the finite size $R_c$ of the WS cell (from $-R_c$ to $+R_c$ along the $z$
direction for slabs, and from 0 to $R_c$ in a circle of radius $R_c$ for rods).
The calculation of the Coulomb energies is also simplified in this case.

The interacting nuclear part ${\cal V}\left(\rho_{n},\rho_{p}\right)$
\cite{BCPM,baldo08,baldo10} consists of two pieces. One piece is directly
related to the EOS of symmetric and neutron matter and depends locally on the
total baryon density as well as on the asymmetry parameter
\begin{equation}
\beta \,=\, \frac{\rho_n \,-\, \rho_p}{\rho_n \,+\, \rho_p} .
\label{eq:beta}
\end{equation}
The other piece of ${\cal V}\left(\rho_{n},\rho_{p}\right)$ is a
finite range term whose contribution to the energy density reads
\begin{eqnarray}
{\epsilon}_{\rm surf}(\rho_{q},\rho_{q'})=
\frac{1}{2}\sum_{q,q'}\rho_{q}(\vec{r})\int \rho_{q'}({\vec{r}\,'}) \, {v}_{q,q'} \, e^{{-(\vec{r} -{\vec{r}\,'})^{2}}/{\alpha^{2}}}d{\vec{r}\,'} ,
\label{eq13}
\end{eqnarray}
which depends on the strengths ${v}_{q,q'}$ of the like and unlike particles and on the range $\alpha$ of the finite range Gaussian form factor. Performing the angular integration, in the spherical case it can be recast as
\begin{eqnarray}
{\epsilon}_{\rm surf}(\rho_{q},\rho_{q'})=
\frac{1}{2}\sum_{q,q'}
\frac{\pi{v}_{q,q'}\alpha^{2}}{r} \rho_{q}(\vec{r})
\int_{0}^{\infty} dr' r' \left[e^{{-(\vec{r} -{\vec{r}\,'})^{2}}/{\alpha^{2}}}
-e^{{-(\vec{r} +{\vec{r}\,'})^{2}}/{\alpha^{2}}}\right]
\rho_{q'}({\vec{r}\,'}) .
\label{eq14}
\end{eqnarray}
It has to be pointed out that in the BCPM energy density functional the
strengths ${v}_{q,q'}$ are chosen in such a way that in the bulk limit (constant
$\rho_{q}$ and $\rho_{q'}$) one recovers the $\rho^{2}$ term of the bulk part of
the energy density (see Ref~\cite{BCPM} for details). For the sake of brevity,
here we avoid writing the detailed expressions of ${\epsilon}_{\rm surf}$ for
the non-spherical shapes.

Once the energy is obtained, the pressure in the inner crust of the neutron star
is computed by the taking the appropriate derivative of the energy with respect
to the size of the WS cell, following e.g. Appendix B of the second reference of
\cite{BSK}.
The result reads as
\begin{equation}
P = P_{\rm g} + P_{\rm free}^{\rm el} + P_{\rm ex}^{\rm el},
\label{eq22}
\end{equation}
which shows that the total pressure in the crust is the sum of the contributions
due to the neutron gas and to the free electrons, plus a corrective term from
electron exchange ($P_{\rm ex}^{\rm el}= \frac{1}{3} {\epsilon}_{\rm ex}^{\rm
el}$).
This will provide the EOS in the crust region of the neutron star. Here we will
not write down the explicit formulae and the numerical method, but we only
describe the results in the next subsection,
leaving a more complete report to a future paper.

\subsection{The crust EOS}
\begin{figure}[t]
%\vskip -10 cm
\begin{center}
\includegraphics[bb= 250 0 400 790,angle=0,scale=0.8]{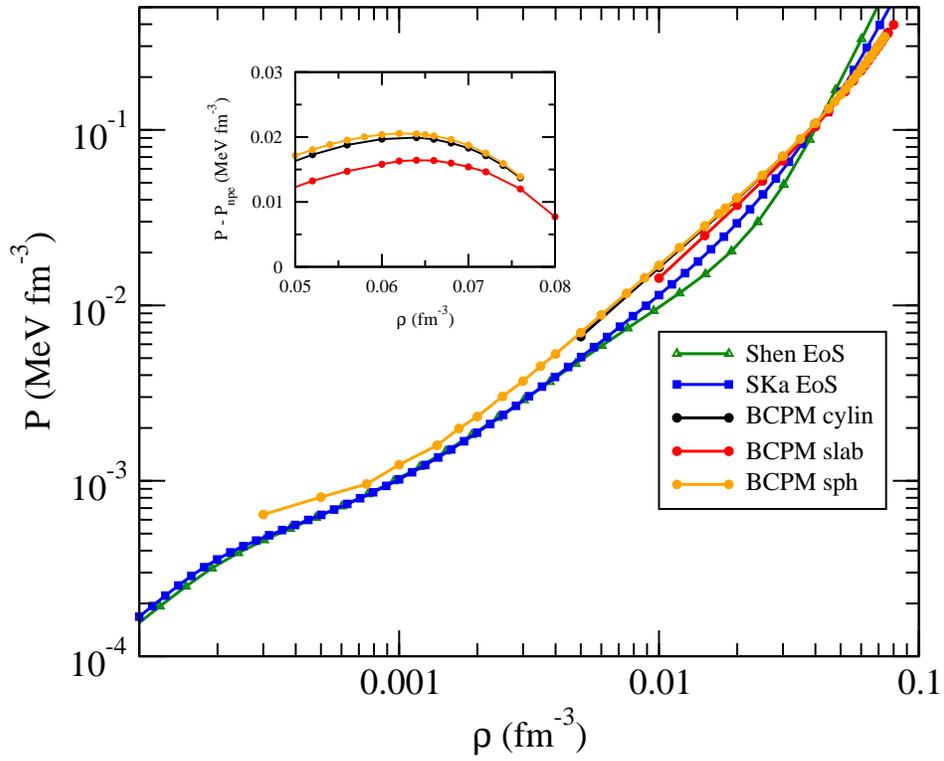}
\vskip -1.5 cm
\caption{Pressure as a function of density in the crust region for the considered Equations of State. For the BCPM functional the pressure corresponding to other possible shapes are also considered. In this case the possible pasta phase seems to have a quite limited extension, see the text for explanations. The lowest density at which the BCPM is reported marks the drip point, essentially for all three EOS. The insert shows the pressure of the BCPM crust EOS at the higher density for the different shapes considered. For convenience at each density the pressure of the corresponding homogeneous matter has been subtracted.}
\label{fig:fig1}\end{center}
\end{figure}

In Fig. \ref{fig:fig1} we compare the EOS of the crust derived from the BCPM
functional \cite{BCPM} with the LS \cite{Lattimer} and SH \cite{Shen} ones.
For BCPM we have not performed a study of the outer crust and we will use the
EOS from ref. \cite{Baym}. In any case, for our purposes the uncertainty on the
outer crust EOS is expected to have a quite limited relevance.
Close to neutron drip, the BCPM results differ slightly from the other
two cases, which agree quite closely between each other. At increasing density
the discrepancy persists but the considered  EOS show some different trends. The
discrepancy for the pressure can be as large as a factor two. The discrepancies
in the crust composition will be reported elsewhere, here we are considering the
overall properties of the NS. For comparison we report the pressure also for the
other structures considered in the BCPM case. The structure of the highest
pressure is the favorable one. One can see that the spherical case seems to
dominate the whole crust, except at the highest density where the other 
shapes can compete, see Fig. \ref{fig:fig1}. Of course other structures are possible \cite{Jicrust}, but
in any case we take this result as an indication that in the BCPM case the pasta
phase can exist only in a very narrow density range and that one can take for
the inner crust EOS the spherical structure. In the EOS of the LS case the pasta
phase seems to be more extended, but we do not perform a detailed comparison, we
just use the  EOS which
includes the most favorable shapes. Despite the crust contains a small fraction
of the NS mass, its EOS and its matching with the core EOS can be relevant for
the determination of the NS radius.
\section{The high-density EOS}
In the center of a NS the baryon density can reach values that are several times larger than the nuclear saturation density. The reason for such a high density is the fact that
a NS is bound by gravity, and it can be kept in hydrostatic
equilibrium only by the pressure produced by the compressed nuclear matter. The core includes most of the NS mass and the high-density EOS is therefore crucial. As mentioned, in our description the core is composed of homogeneous asymmetric nuclear matter and of a gas of
electrons and eventually muons. The asymmetry is fixed by the beta equilibrium,
which in turn is determined mainly by the symmetry energy as a function of
density \cite{Lattimer2004}. Phenomenologically, the symmetry energy can be extracted only below the saturation density \cite{Danie}, and therefore it is challenging to predict the
symmetry energy at high density, as demanded by NS studies. The EOS inside the
NS can be strongly influenced by the value of the asymmetry, since the stiffness of the EOS depends on the nuclear matter composition. \par
\begin{figure}[t]
%\vskip -10 cm
\begin{center}
\includegraphics[bb= 0 0 300 790,angle=-90,scale=0.7]{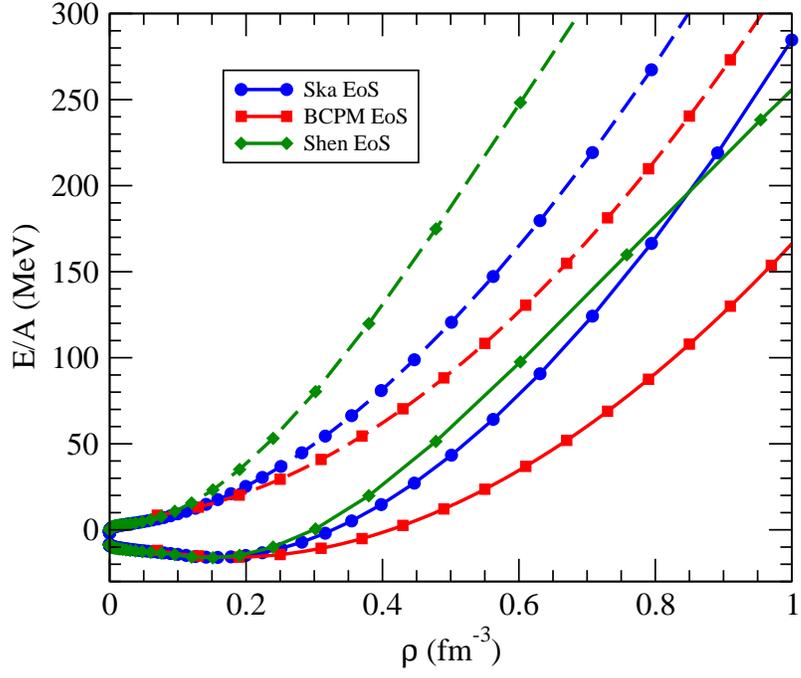}
\vskip 6.5 cm
\caption{Equations of State for pure neutron matter (dashed
lines) and symmetric nuclear matter (solid lines), for the three considered
cases.}
\label{fig:fig2}\end{center}
\end{figure}

It is known that the dependence of the EOS on the asymmetry parameter $\beta$ is almost exactly quadratic. In this case,
the symmetry energy is just the difference between the EOS of pure neutron matter and the EOS of symmetric matter. We report in Fig. \ref{fig:fig2} the neutron and symmetric
matter EOS for the cases we are considering.
Just above the saturation density the differences among the EOS are drastic,
especially for the pure neutron matter. Just on inspection, it is also apparent
that the symmetry energy at higher density is quite different. For symmetric
nuclear matter the different values of the incompressibility are 
responsible for the discrepancies. Indeed, the incompressibility is $K \,=\,
213$ MeV for the BCPM EOS,
$264$ MeV for the LS EOS in its Ska version, and $274$ MeV for the SH EOS.
The symmetry energy at saturation has comparable values for all EOS, with a slightly higher value for the SH EOS, while there are large discrepancies at higher density, where the LS EOS
shows the smallest value among the three.
\par
\begin{figure}[t]
%\vskip -10 cm
\begin{center}
\includegraphics[bb= 0 0 300 790,angle=-90,scale=0.7]{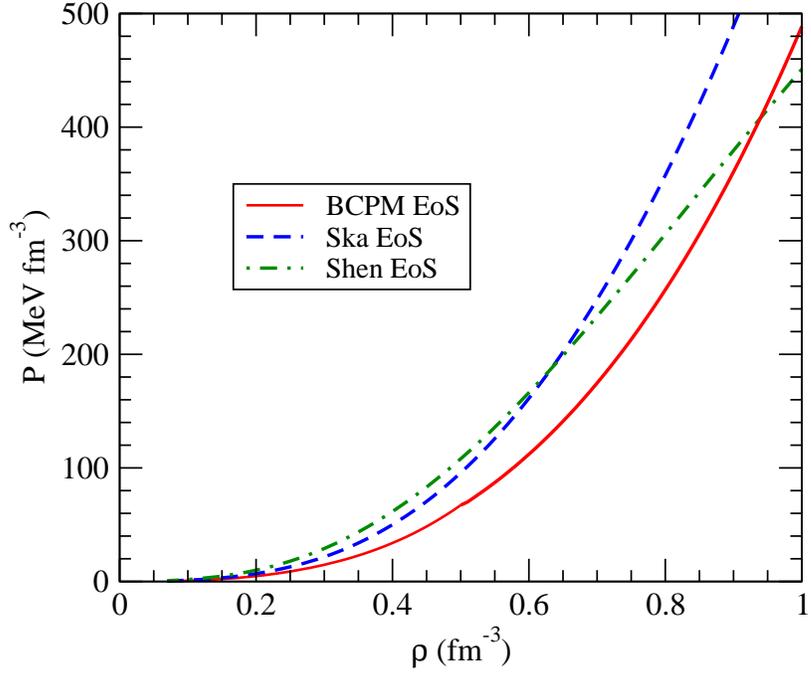}
\vskip 6.5 cm
\caption{Equation of State for beta-stable matter for the three considered cases.}
\label{fig:fig3}\end{center}
\end{figure}
\begin{figure}[t]
%\vskip -10 cm
\begin{center}
\includegraphics[bb= 0 0 300 790,angle=-90,scale=0.6]{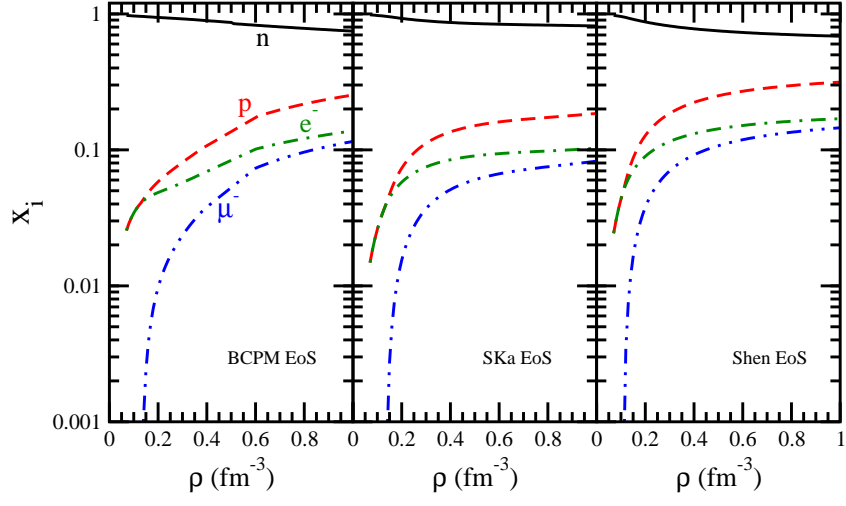}
\vskip 2.5 cm
\caption{Composition of Neutron Star matter as a function of baryon density. The lowest (dash dot dot) line corresponds to the muon fraction.}
\label{fig:fig4}\end{center}
\end{figure}

For each one of the EOS we have calculated the beta equilibrium and the corresponding asymmetric matter EOS, including leptons, which is the relevant quantity we need for the NS core. They are reported in Fig. \ref{fig:fig3}.
One can see that now the discrepancies among the EOS tend to be reduced. This is a consequence of a partial compensation between the value of the symmetry energy and the effect of the composition.
In fact, if the symmetry energy is larger, then a larger fraction of protons is present. This tends to reduce the stiffness of the resulting EOS for matter in beta equilibrium. This can be seen from  Fig. \ref{fig:fig4}, where the NS matter composition is reported for the different EOS. The proton fraction goes in parallel with the density dependence of the symmetry energy.
However, relevant discrepancies among the EOS still persist, which will be reflected in the corresponding NS structure.
\par
\section{Results}
\begin{figure}[t]
%\vskip -10 cm
\begin{center}
\includegraphics[bb= 0 0 300 790,angle=-90,scale=0.8]{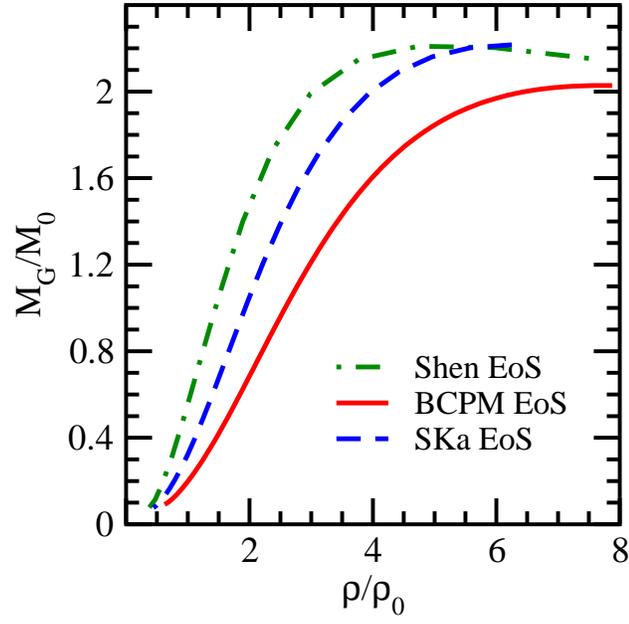}
\vskip 7. cm
\caption{Neutron Star gravitational mass as a function of
central baryon density for the three considered cases.}
\label{fig:fig5}\end{center}
\end{figure}

Once the EOS of the nuclear matter which is present throughout the NS is known, one can use the celebrated Tolman-Oppenheimer-Volkoff \cite{Shap} equations for spherically symmetric NS:
\begin{eqnarray}
{d P\over d r}&=&- G\, {\epsilon m \over r^2} \left(1 + {P \over \epsilon} \right) \left(1 + {4\pi P r^3\over m } \right) \left(1 - {2 G m \over r }\right)^{-1} \nonumber \\
{d m\over d r}&=&4\pi r^2 \epsilon \,,
\label{eq:OV} 
\end{eqnarray}
\noindent where $G$ is the gravitational constant, $P$ the pressure, $\epsilon$ the energy density,  $m$ the mass enclosed within a radius $r$, and $r$ the (relativistic) radius coordinate. To close the equations we need the relation between pressure and density, $P \,
=\, P(\epsilon)$, i.e. just the EOS. Integrating these equations one gets the mass and radius of the star for each central density. Typical values are
1--2 solar masses ($M_\odot$) and about 10 km, respectively. This indicates the
extremely high density of the object. It turns out that the mass of the NS has a
maximum value as a function of radius (or central density), above which the star
is unstable against collapse to a black hole. The value of the maximum mass
depends on the nuclear EOS, so that the observation of a mass higher than the
maximum mass allowed by a given EOS simply rules out that EOS.
 The considered EOS are compatible with the largest mass observed up to now,
which is close to $2.01 M_{\odot} \pm 0.04$  \cite{Anton}, as it can be seen in
Fig. \ref{fig:fig5}, where the relation between mass and central density is
reported. Even if it can look unlikely that this value is indeed the largest
possible NS mass, we take this result as an indication of the validity of the
EOS.
 As expected, when the incompressibility increases the NS central density
decreases for a given mass.
\par
\begin{figure}[t]
%\vskip -10 cm
\begin{center}
\includegraphics[bb= 0 0 300 790,angle=-90,scale=0.7]{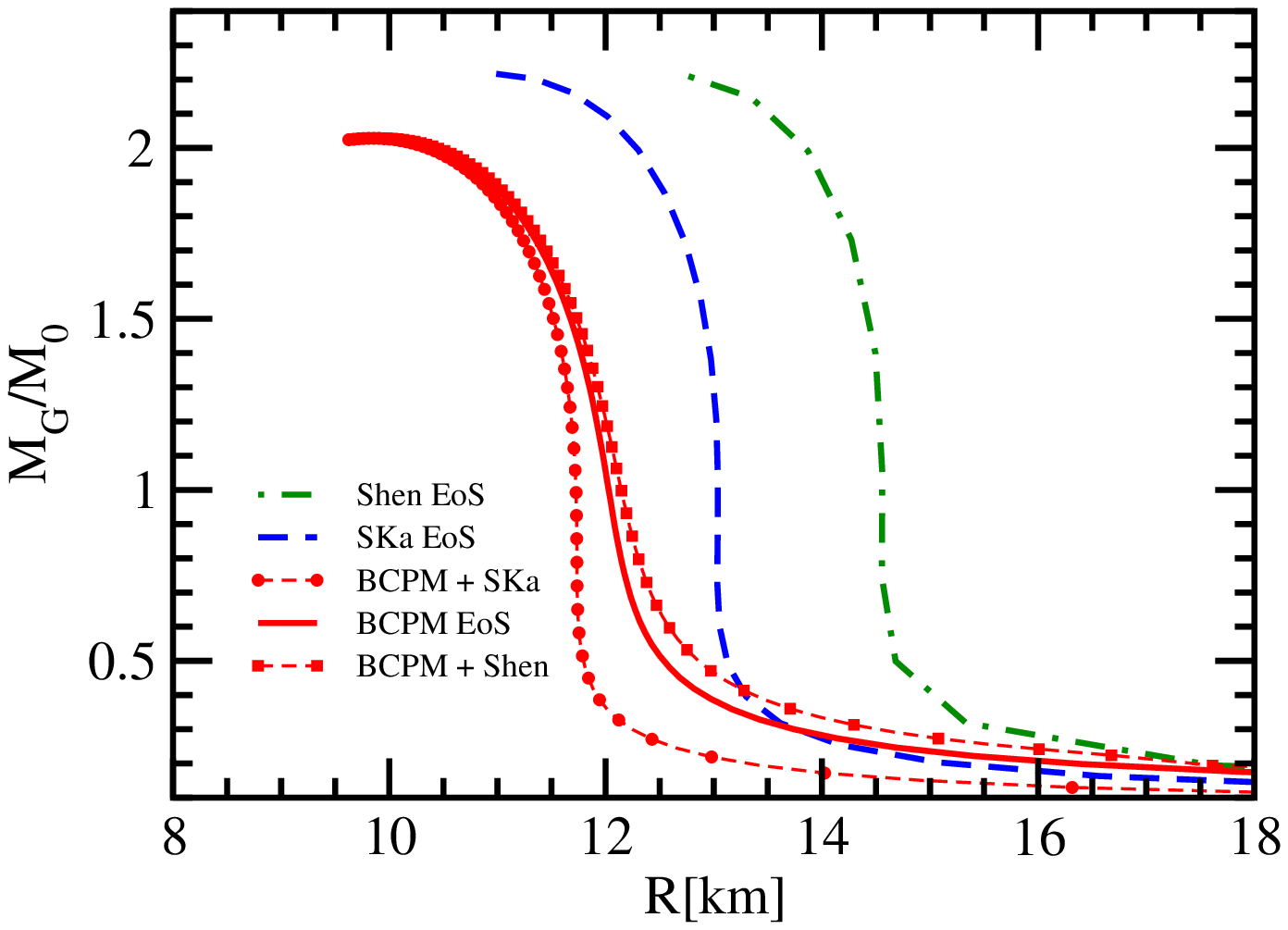}
\vskip 6. cm
\caption{Neutron Star gravitational mass as a function of the
radius. The full line corresponds to the BHF Equation of State, together with
the corresponding BCPM functional, see the text for details. The squares
correspond to the BHF EOS for the core and the Shen EOS \cite{Shen,SHweb} for the
crust. The circles correspond to the BHF EOS for the core and the Lattimer and
Swesty EOS \cite{Lattimer,LSweb} for the crust. See the text for details.}
\label{fig:fig6}\end{center}
\end{figure}

We will now discuss the relevance of the crust EOS on the NS radius, which is
the main point of our work. 
We have calculated the mass-radius relationship for the three 
considered EOS, which extend to the whole NS structure from the crust to the inner core.
 The results are reported
Fig. \ref{fig:fig6}. As one can expect, the value
of the radius increases with the NS matter incompressibility, the largest one
corresponding to the highest incompressibility, if the comparison is restricted
up to about 0.5 fm$^{-3}$.
The latter is the maximum density that is reached at the center of the NS for the largest mass obtained with the SH EOS. As one can see from Fig. \ref{fig:fig3}, at higher density the pressure of the SH EOS becomes smaller than the LS one, but those density values are not reached in stable NS for the SH EOS.
\par  
In order to study the role of the crust in the values of the NS radius we have performed NS calculations by using different crust EOS but fixing the
same EOS for the core NS matter. Specifically, we have performed calculations 
using the BHF EOS for the core and joining it with the crust EOS from LS and SH
models. The results are reported in the same Fig. \ref{fig:fig6} (circles and
squares, respectively). One can see that the use of SH crust EOS makes only a
relatively small effect on the mass-radius relationship, while a larger effect
is observed for the LS crust EOS. Of course, the largest effect is obtained for
the smaller masses, where the crust has a prominent relevance. However, even for
masses up to 1.2--1.3 solar masses, the variations of the radius are quite
sizeable. This also shows the relevance of using the EOS derived from the same
theoretical scheme both for the crust and for the core. The matching of the core
EOS with a crust EOS from a different model can introduce further uncertainties
for the NS structure. \par
Finally we found that
the mass-radius relationship calculated in 
ref. \cite{Haensel} turns out to be 
surprisingly close to the one obtained
with BHF EOS and the consistent
BCPM functional for the crust.
Maybe this can be explained by the
close value of the bulk incompressibility,
which is 230 MeV, in comparison
with the value 223 MeV for BHF.
However the composition in the bulk
is quite different in the two cases,   
and furthermore the crust EOS are 
appreciably different. Therefore 
the agreement must be considered
at least partially fortuitous.    

\section{Conclusions}
\label{s:end}
 We have presented a study of Neutron Star structure by focusing on models which are able to describe on the same physical framework both the core and the crust regions. To this purpose we have
performed calculations of NS structure based on the Equation of State derived
within the Brueckner-Hartree-Fock scheme. The EOS is the basis of an accurate
Energy Density functional that was recently \cite{BCPM} devised to reproduce
nuclear binding energies throughout the nuclear mass table. As such,
it was used to calculate the structure of NS crust, within the Thomas-Fermi approach. This EOS was then the basis of a microscopic approach of the overall NS structure, from the crust to the core.
To our knowledge this is the first microscopic calculation that embraces the whole density range that
appears inside a NS. The results were compared with other more phenomenological models that also are able to describe the whole NS structure. To emphasize the role of the NS crust, we have performed calculations by employing the microscopic BHF EOS in the NS core and joining it to the crust EOS
from the more phenomenological approaches. The results indicate that the particular EOS crust that is employed can produce a sizeable effect on the NS radius, not only for the lightest NS but also for standard NS masses. This indicates the relevance of using models that are based on the same physical framework both for the crust and the core.              

\acknowledgments
M.C., B.K.S., and X.V. acknowledge the support from the Consolider Ingenio 2010
Programme CPAN CSD2007-00042, from Grant No. FIS2011-24154 from MICINN and
FEDER, and from Grant No. 2009SGR-1289 from Generalitat de Catalunya. B.K.S. greatly acknowledges the financial support from Grant No. CPAN10-PD13 from CPAN (Spain).

%%%%%%%%%%%%%%%%%%%%%%%%%%%%%%%%%%%%%%%%%%%%%%%%%%%%%%%%%%%%%%%%%%%%%%%%%%%%%%%%

\vfill\eject

\vfill\eject
\vfill\eject
\vfill\eject
\vfill\eject
\vfill\eject


\begin{thebibliography}{99}
\bibitem{BCPM} M. Baldo, L. Robledo, P. Schuck and X. Vinas, Phys. Rev. {\bf C87}, 064305 (2013).
\bibitem{baldo08}
M. Baldo, P. Schuck and X. Vi\~nas, Phys. Lett. B {\bf 663}, 390 (2008).
\bibitem{baldo10}
M. Baldo, L. M. Robledo, P. Schuck and X. Vi\~nas, J. Phys. G {\bf 37}, 064015 (2010).
\bibitem{book}
% THE MANY-BODY THEORY OF THE NUCLEAR EQUATION OF STATE
 M. Baldo,
 {\em Nuclear Methods and the Nuclear Equation of State},
 International Review of Nuclear Physics, Vol. 8
 (World Scientific, Singapore, 1999).
\bibitem{song}
% BETHE-BRUECKNER-GOLDSTONE EXPANSION IN NUCLEAR MATTER
 H. Q. Song, M. Baldo, G. Giansiracusa, and U. Lombardo,
 Phys. Rev. Lett. {\bf 81}, 1584 (1998);
% BETHE-BRUECKNER-GOLDSTONE EXPANSION IN NEUTRON MATTER
 M. Baldo, G. Giansiracusa, U. Lombardo, and H. Q. Song,
 Phys. Lett. {\bf B473}, 1 (2000);
% HIGH DENSITY SYMMETRIC NUCLEAR MATTER IN THE BBG APPROACH
 M. Baldo, A. Fiasconaro, H. Q. Song, G. Giansiracusa, and U. Lombardo,
 Phys. Rev. {\bf C65}, 017303 (2001).
 \bibitem{uix}
% THREE-NUCLEON INTERACTION IN 3-, 4- AND INFINITE-BODY SYSTEMS
 J. Carlson, V. R. Pandharipande, and R. B. Wiringa,
 Nucl. Phys. {\bf A401}, 59 (1983);
% MOMENTUM DISTRIBUTIONS IN A = 3 AND 4 NUCLEI
 R. Schiavilla, V. R. Pandharipande, and R. B. Wiringa,
 Nucl. Phys. {\bf A449}, 219 (1986);
%% QUANTUM MONTE-CARLO CALCULATIONS OF A<=6 NUCLEI
% B. S. Pudliner, V. R. Pandharipande, J. Carlson and R. B. Wiringa,
% Phys. Rev. Lett. {\bf 74}, 4396 (1995);
% QUANTUM MONTE CARLO CALCULATIONS FOR NUCLEI WITH A<=7
 B. S. Pudliner, V. R. Pandharipande, J. Carlson, S. C. Pieper, and R. B. Wiringa,
 Phys. Rev. {\bf C56}, 1720 (1997).
\bibitem{bbb}
% MICROSCOPIC NUCLEAR EOS WITH THREE-BODY FORCES AND NEUTRON STAR STRUCTURE
 M. Baldo, I. Bombaci, and G. F. Burgio,
 Astron. Astrophys. {\bf 328}, 274 (1997).
\bibitem{zhou}
% THREE-BODY FORCES AND NEUTRON STAR STRUCTURE
 X. R. Zhou, G. F. Burgio, U. Lombardo, H.-J. Schulze, and W. Zuo,
 Phys. Rev. {\bf C69}, 018801 (2004).
 \bibitem{Lattimer} J.M. Lattimer and F.D. Swesty, Nucl. Phys. {\bf A535}, 331 (1991).
\bibitem{Shen} H. Shen, H. Toki, K. Oyamatsu and K. Sumiyoshi, Prog. Theor. Phys.  {\bf 100}, 1013 (1998); H. Shen, H. Toki, K. Oyamatsu and K. Sumiyoshi, Nucl. Phys.  {\bf A637}, 435 (1998).
\bibitem{LSweb} http://www.astro.sunysb.edu/lattimer/EOS/main.html 
\bibitem{SHweb} http://user.numazu-ct.ac.jp/~sumi/eos/ 
\bibitem{Haensel} F. Douchin and P. Haensel, Astron. Astrophys. {\bf 389}, 151 (2001).
\bibitem{Sly} E. Chabanat, P. Bonche, P. Haensel, J. Meyer and R. Schaeffer, Nucl. Phys. {A635}, 231 (1998).
\bibitem{FP} B. Friedman and V.R. Pandharipande, Nucl. Phys. {\bf A361}, 502 (1981).
%
\bibitem{BSK} A.F. Fantina, N. Chamel, J.M. Pearson, and S. Goriely,
Astron. Astrophys. {\bf 559}, A128 (2013);
J. M. Pearson, N. Chamel, S. Goriely, and C. Ducoin,
Phys. Rev. C {\bf 85}, 065803 (2012).
%
\bibitem{goriely10} S. Goriely, N. Chamel, and J.M. Pearson Phys. Rev. {\bf C82}, 035804 (2010).
%
\bibitem{potekhin13}
A.Y. Potekhin, A.F. Fantina, N. Chamel, J.M. Pearson, and S. Goriely,
Astron. Astrophys. {\bf 560}, A48 (2013).
%
\bibitem{sil02}
Tapas Sil, J. N. De, S. K. Samaddar, X. Vi\~nas, M. Centelles, B. K. Agrawal, and S. K. Patra,
Phys. Rev. C {\bf 66}, 045803 (2002).
%
%\bibitem{ETF} M. Centelles, M. Del Estal and X. Vinas, Nucl. Phys. A635, 193 (1998).
%
%
\bibitem{oyamatsu93}
K. Oyamatsu, Nucl. Phys. A {\bf 561}, 431 (1993).
\bibitem{onsi08}
M. Onsi, A. K. Dutta, H. Chatri, S. Goriely, N. Chamel, and J. M. Pearson,
Phys. Rev. C {\bf 77}, 065805 (2008).
%
\bibitem{gogelein07}
P. G\"{o}gelein and H. M\"{u}ther, Phys. Rev. C {\bf 76}, 024312 (2007).
\bibitem{cheng97}
K. S. Cheng, C. C. Yao, and Z. G. Dai, Phys. Rev. C {\bf 55}, 2092 (1997).
\bibitem{avancini08}
S. S. Avancini, D. P. Menezes, M. D. Alloy, J. R. Marinelli, M. M. W. Moraes, and C. Provid\^{e}ncia, 
Phys. Rev. C {\bf 78}, 015802 (2008).
\bibitem{avancini09}
S. S. Avancini, L. Brito, J. R. Marinelli, D. P. Menezes, M. M. W. de Moraes, C. Provid\^{e}ncia, and A. M. Santos,
Phys. Rev. C {\bf 79}, 035804 (2009).
\bibitem{avancini12}
F. Grill, C. Provid\^{e}ncia, and S. S. Avancini, Phys. Rev. C {\bf 85}, 055808 (2012).
\bibitem{Baym} G. Baym, C. Pethick and P. Sutherland, Astrophys. J. {\bf 170}, 299 (1971).
\bibitem{Jicrust} H. Pais and J.R. Stone, Phys. Rev. Lett. {\bf 109} 151101 (2012). 
\bibitem{Lattimer2004} J. M. Lattimer and M. Prakash, Science {\bf 304}, 536 (2004).
\bibitem{Danie} M. B. Tsang, J. R. Stone, F. Camera, P. Danielewicz, S. Gandolfi, K. Hebeler,
C. J. Horowitz, Jenny Lee, W. G. Lynch, Z. Kohley, R. Lemmon, P. M$\rm \ddot oller$, 
T. Marukami, S. Riordan, X. Roca-Maza, F. Sammarruca, A. W. Steiner, I. Vida$\rm\tilde{n}$a, and 
S. J. Yennello, Phys. Rev. C{\bf 86}, 015803 (2012); 
M. B. Tsang, C. K. Gelbke, X. D. Liu, W. G. Lynch, W. P. Tan, G. Verde, H. S. Xu, W. A. Friedman, R. Donangelo, S. R. Souza, C. B. Das, S. D. Gupta, and D. Zhabinsky, 
Phys. Rev. C{\bf 64}, 054615 (2001);
T. X. Liu, W. G. Lynch, M. B. Tsang, X. D. Liu, R. Shomin, W. P. Tan, G. Verde, A. Wagner, H. F. Xi, H. S. Xu, B. Davin, Y. Larochelle, R. T. de Souza, R. J. Charity, and L. G. Sobotka, 
Phys. Rev. C{\bf 76}, 034603 (2007).  
\bibitem{Shap} S.L. Shapiro and S.A. Teukolski {Black Holes, White Dwarfs and Neutron Stars},
(John Wiley and Sons, New York, 1983). 
%\bibitem{Demo} P.B. Demorest, T. Pennucci, S.M. Ransom, M.S.E. Roberts and J.W.T. Hessels, %Nature(London){\bf 467} 1081 (2010).
\bibitem{Anton} J. Antoniadis  {\it et al.}, Science {\bf 340}, 6131 (2013).




\end{thebibliography}
\end{document}